\documentclass[10pt]{article}
\usepackage{graphicx,color} 
\usepackage{amsmath,mathrsfs}
\usepackage{natbib}

\usepackage[english]{babel}
\usepackage[T1]{fontenc}

\newcommand{\at}{{\char '100}}

\newcommand{\T}[1]{#1^{\mathnormal{\scriptscriptstyle T}}} 

\renewcommand{\theenumi}{(\kern -0.15ex{\roman{enumi}})}

\begin{document}
\onecolumn

\title{Symmetry: a bridge between nature and culture \\
Simmetria: un ponte tra cultura e natura\\
Sym\'etries : un pont entre nature et culture}

\author{Amaury Mouchet\\ \small
Laboratoire de Math\'ematiques
  et de Physique Th\'eorique, \\ \small
Universit\'e Fran\c{c}ois Rabelais de Tours --- \textsc{\textsc{cnrs (umr 7350)}},\\ \small
 F\'ed\'eration Denis Poisson,\\ \small
 Parc de Grandmont 37200  Tours,  France\\ \small
 mouchet\at phys.univ-tours.fr\\ \small
\date{March, 28, 2014}
}

\maketitle

This text, to be published in the volume \textit{Imagine Math 4} (UMI and Istituto Veneto), is
the proceeding of a talk given at the conference \textit{Mathematica e
  cultura 2014} organized by Michele Emmer at the \textit{Istituto
  Veneto di Scienze, Lettere e Arti, Palazzo Franchetti} in March
2014: \\ \texttt{\small
  www1.mat.uniroma1.it/ricerca/convegni/Venezia/2014/}\\ The
corresponding on line video is available at\\ \texttt{\small
  www.youtube.com/playlist?list=PLfcFPNXyAOqatuGE3E9ZqwVqt10ieI3Jy
}

\begin{quotation} The possibility of  translation implies the existence of an
   invariant.  To translate is precisely to disengage this
  invariant.[\dots]\\ 
  \indent What is  objective must be common to many
  minds and consequently transmissible from one to the other, and this
  transmission can only come about by [a] ``discourse'' [\dots] we are
  even forced to conclude: no discourse, no  objectivity.
  [\dots]\\ 
  \indent Now what is science? I have explained [above], it
  is before all a classification, a manner of bringing together facts
  which appearances separate, though they were bound together by some
  natural and hidden kinship. Science, in other words, is a system of
  relations.  Now we have just said, it is in the relations alone that
  objectivity must be sought; it would be vain to seek it in beings
  considered as isolated from one another.\\ 
  \indent To say that
  science cannot have objective value since it teaches us only
  relations, is to reason backwards, since, precisely, it is
  relations alone which can be regarded as objective.\\ 
  \indent External
  objects, for instance, for which the word \emph{object} was invented,
  are really \emph{objects} and not fleeting and fugitive appearances,
  because they are not only groups of sensations, but groups cemented
  by a constant bond. It is this bond, and this bond alone, which is
  the object in itself, and this bond is a relation.\\
\citep[\S\S 4 and 6]{Poincare02a}
\end{quotation}

Are symmetries discovered or rather invented by humans ? The stand you
may take firmly here reveals a lot of your epistemological
position. Conversely, the arguments you may forge for answering to
this question, or to one of its numerous narrower or broader
variations, shape your whole philosophical thoughts; not
specifically about science, by the way.  I will
try to show how physics helps to (re)consider this issue.  Indeed,
in the \textsc{xx}th century, physicists have not only extended the
notion of symmetry much beyond the rich heritage of geometers but they
have also deeply rooted it into the natural world. As a consequence,
we can foresee that nature and culture are so coherently entangled one
with the other that the two possible answers of what seems an
inescapable alternative appear to be two banks continuously connected
by one single bridge.

After some brief recalls in~\S~\ref{sec:transformation_invariance}
about the articulation between two essential facets of symmetry,
namely the notion of transformation and the notion of invariance, I
will quickly review in~\S~\ref{sec:physicaltransformations} what kind
of transformations physicists talk about. Then,
in~\S~\ref{sec:physicallaw}, I will illustrate,
with one of the simplest examples, what is meant by the
transformation of a physical law. This
will allow us to understand how invariance is indeed a \textit{raison
  d'\^etre} of the science laws themselves.  In the next section,
\S~\ref{sec:noether}, I will explain how the mathematical work of
Emmy Noether and its repercussions in physics has strengthened even
more the bind between invariances, that actually make science possible,
and the local conservation laws, from which the physical fundamental
objects (the quantum particles) come to existence.  Eventually, I will
conclude in~\S~\ref{sec:philosophy} by some remarks that go beyond
physics and concern more generally rational thinking.
 
\section{Transformation and invariance}\label{sec:transformation_invariance}

The modern concept of symmetry has many 
facets\footnote{Please, see~\citep{Mouchet13b}  for a more technical paper on the subject where four facets are distinguished and extensively discussed. 
For a broader audience (in French) see \citep{Mouchet13a}. } 
and we shall focus
here on two of them only: the notion of \emph{transformation}
and the notion of \emph{invariance}. 
 
We shall always require that the set of the transformations we are
considering constitutes a \emph{group}~$\mathsf{G}$ i.e.  a set whose any
element~$T$ is a one to one mapping. Besides, the composition of two
elements of~$\mathsf{G}$ remains an admissible transformation, that is
an element of~$\mathsf{G}$.  Then, any transformation can be undone
and the identity (``doing nothing'') is a somehow trivial
transformation that belongs to~$\mathsf{G}$.

\begin{figure}[!Ht]\begin{center}
\includegraphics[width=8cm]{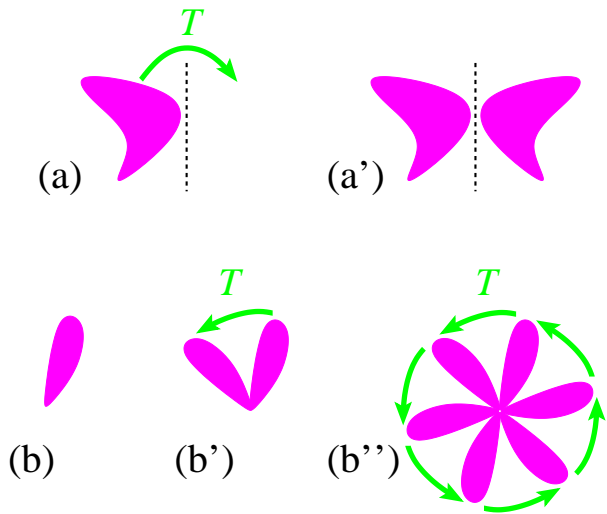}\caption{\label{fig:petales} 
Above, the group of transformations~$\mathsf{G}_a$ is just 
the doublet~$\{T,T^2=1\}$ where~$1$
denotes the identity and~$T$ the mirror symmetry with respect to the dashed line. The pairing of the object~(a) and its image by~$T$ 
constitute a \textit{symmetrical} object 
(a'), that is an object globally invariant by~$T$ even if each of its two parts is generally not invariant. 
Below, the group~$\mathsf{G}_b$  is made of a rotation of 1/6th 
turn together with its 5 distinct repetitions~$\{T,T^2,T^3,T^4,T^5,T^6=1\}$.
One petal of the flower~(b) is not 
invariant by any rotation in~$\mathsf{G}$ but the flower~(b'') made of 6 petals is.
}
\end{center}
\end{figure}

The connexion between transformations and invariance is rather
straightforward. If you take an object on which the transformations
in~$\mathsf{G}$ can act, and if you collect all the images obtained by
applying all the elements of~$\mathsf{G}$ you will built an object, so
called a ``symmetrical object'', that remains, by construction,
unchanged if you apply on it any transformation in the group. The
latter is known as the \emph{symmetry group of the (symmetrical)
  object}. Of course, one can follow the reverse path: saying that an
object is symmetrical under the group~$\mathsf{G}$ means that it can
be reduced in elementary parts differing one from the other by a
transformation in~$\mathsf{G}$.  The figure~\ref{fig:petales}
provides two simple illustrations.

\section{Physical transformations}\label{sec:physicaltransformations}

The interplay between transformations and invariance explained in the
previous section goes back even before the name of symmetry was forged
from the ancient Greek $\sigma\acute\upsilon\mu$ (with, concordance,
harmony) $\mu\acute\epsilon\tau\rho o \nu$ (measure, proportion).
This interplay is ubiquitous in nature and art, specially when some
parsimony is required (the elaboration of 2 lungs, 6 petals, 12
pillars, 1000~golden mosaic tiles or honey cells, etc.). Even in music and
literature, specially for rhythm or rhymed verses,
 symmetry may help the memory
of the bard.

Geometrical transformations like mirror symmetry, rotations or translations
 are privileged by intuition and actually play a crucial role in our representation of space. 
\begin{quotation}It is seen that experiment plays a considerable role in the genesis of
geometry; but it would be a mistake to conclude from that that
geometry is, even in part, an experimental science. If it were
experimental, it would only be approximative and provisory. And what a
rough approximation it would be! Geometry would be only the study of
the movements of solid bodies; but, in reality, it is not concerned
with natural solids: its object is certain ideal solids, absolutely
invariable, which are but a greatly simplified and very remote image
of them. The concept of these ideal bodies is entirely mental, and
experiment is but the opportunity which enables us to reach the
idea. The object of geometry is the study of a particular "group"; but
the general concept of group pre-exists in our minds, at least
potentially. It is imposed on us not as a form of our sensitiveness,
but as a form of our understanding; only, from among all possible
groups, we must choose one that will be the standard, so to speak, to
which we shall refer natural phenomena.\\
Experiment guides us in this choice, which it does not impose on us. 
It tells us not
what is  truest, but what is the most convenient geometry. 
It will be noticed that my
description of these fantastic worlds 
has required no language other than that of
ordinary geometry. Then, were we transported to those worlds, there would be no
need to change that language. Beings educated there would no doubt find it more
convenient to create a geometry different from ours, and better adapted to their
impressions; but as for us, in the presence of the same impressions, it is certain that
we should not find it more convenient to make a change.\\
\citep[Conclusions]{Poincare1895a}
\end{quotation}
In fact, with 
Galileo Galilei's famous arguments on the invariance of experiments when embarked at constant
speed on a boat, since the very beginning of modern physics,
transformations are considered that are not pure  (static) 
geometric displacements in the three dimensional Euclidian space. 
However, in the \textsc{xx}th century, physicists have
considered far more general transformations and have worked out much more
abstracted geometries in ``fantastic worlds'':\\[.5\baselineskip]
(i) In the theory of Relativity one builds up a 4-dimensional geometry of
space-time that comes with dynamical transformations where time and
space coordinates are blended together through linear transformations (the Poincar\'e group in Special Relativity) 
or any smooth transformation (the group of space-time
diffeormorpisms in General Relativity).\\[.5\baselineskip]
\\ (ii) In quantum theory,
transformations act on very abstract spaces made up of algebraic
objects --- like wavefunctions or operators --- that represent
quantum states. In such spaces, a rotation of one turn may have some
significant effects (typically on half-integer spins). Transformations
can also be made local (the so-called gauge transformations); for
instance, a rotation whose angle depends on when and where it is done.\\[.5\baselineskip]
\\ (iii) Transformations may involve the exchange of quantum
particles. For instance, one wish to compare the stability of two
atomic nuclei, the second being obtained from the first by replacing
protons by neutrons and vice-versa. Considering these kinds of
permutations is crucial if one wants to understand the existence of
antiparticles (very much like in the mirror symmetry in
figure~\ref{fig:petales} that provides a simple rule for constructing
the image of an object, the change of the electric charge, grossly
speaking, may transform a particle into its antiparticle and
conversely) or collective effects (electron shells, laser light,
superconductivity, superfluidity, etc.) where quantum particles are
indistinguishable in a way that one cannot conceive in our macroscopic
world.

\section{Transformation of a physical law}\label{sec:physicallaw}

The transformations of some physical systems or of their
observers (a rotation of a planet,
a boost of a boat, a free fall of a lift, the 
substitution  of an electron by a positron) come along with a possible transformation of the physical laws themselves. Take for instance 
the relation
\begin{equation}\label{eq:pendulum}
  P=2\pi\sqrt{\frac{\ell}{g}}
\end{equation}
between the period of oscillations~$P$ of a pendulum, its length~$\ell$ and
the acceleration of gravity~$g$. In terms of transformations, this formula
tells you that if you change the length into~$\T{\ell}=4\ell$, the period
will transform according to~$\T{P}=2P$. 
In other words, the relationship between the transformed
 quantities~$(\T{\ell}=4\ell,\T{P}=2P)$ remains the same, namely
\begin{equation}
  \T{P}=2\pi\sqrt{\frac{\T{\ell}}{g}}\;.
\end{equation}
In fact,
the very existence of such a law is a manifestation of some
invariance under
some transformations. This formula is independent of 
the date and location of the observation: it remains correct in Venice
 in 2014 as well as in Pisa in 1583. Even more, it is still exact on the Moon
provided we take into account the transformation of~$g_{\mathrm{Earth}}$ 
into~$g_{\mathrm{Moon}}\simeq g_{\mathrm{Earth}}/6$. 

But it is however crucial
to remember that a physical law must not be reduced to
a formula like~\eqref{eq:pendulum} alone; it must also 
come with some domain of validity and, whenever it is possible, a
 quantitative estimate of the unavoidable uncertainties.
We have no need to evoke the ``revolutions'' of  Relativity 
or  quantum physics
to rule out a formula like~\eqref{eq:pendulum}; even when remaining
within classical Newtonian mechanics, it becomes inaccurate
 as soon as  we take large
variations of the amplitude of the oscillations, or increase the viscosity
of  the ambient
medium. These transformations require a transformation of the equation,
since neither amplitudes nor the viscosity appear in~\eqref{eq:pendulum}.
It is only when it comes with a domain of validity that a science law
may acquire an indelible status; all the more that a more general model or
theory helps to delimit its range.  
Following
Z\'enon, the Renaissance protagonist of Marguerite Yourcenar's novel
\textit{The Abyss},
\begin{quotation}\small{
I have refrained from making an idol of truth, 
preferring to leave to it its more modest name of exactitude} 
\cite[p.~123, A conversation in Innsbruck]{Yourcenar76a}.
\end{quotation}

In fact, the whole science consists precisely in filtering from an
uncountable set of parameters the very few ones (the relevant
variables) that may influence the dynamics of a given system and that
allow to make reasonable predictions.  If we want, say, to obtain two
oscillations in the free air during one second with an amplitude of $5^\circ$ with a
precision up to $10\%$, formula~\eqref{eq:pendulum} is
sufficient. Even if we take a more precise and therefore more
elaborated formula (that takes into account the shape of the pendulum,
its amplitude and the ambient viscosity), the mass of the observer, the
intensity of the surrounding light, the socio-economic structure
that supports the cost of the experiment, the  
position of the moons of Jupiter still
remain irrelevant variables; in other words, the formula is
invariant with respect to any realistic transformation of the 
latter parameters.

The two facets of symmetry recalled in
section~\ref{sec:transformation_invariance} lie at the source of
the \emph{universality} of science laws (including their domain of validity)
 through two angular stones: the invariance
with respect to the transformation of the system (\emph{reproducibility}) and
the invariance with respect to the transformation of the observer
(\emph{objectivity}).

\section{Symmetry and conservation laws}\label{sec:noether}

Motivated by the newborn theory of General Relativity, the
mathematician Emmy Noether published in 1918 extremely profound
results that connect any general (global) invariance to the special
(local) invariance with respect to time translation \citep{KosmannSchwarzbach10a}.  More precisely
the main Noether's theorem stipulates that \textit{for any continuous
  group of transformations (Lie group) under which an optimization
  problem is invariant, there exists a conserved quantity}.  The
continuous groups studied by Sophus Lie are made of transformations labeled
by parameters that can vary continuously. For instance, unlike the
discrete groups illustrated in figure~\ref{fig:petales}, if we
consider every possible angle of rotation (not only the 6 ones
of~$\mathsf{G}_b$), we recover the continuous group of rotations in
the plane whose continuous variable is precisely given by the angle of
rotation.  By an optimization problem (also known as a variational
problem), it is meant that the issue can be formulated by saying that
its solutions maximize or minimize some global quantity\footnote{More precisely,
corresponds to a stationnary value of this quantity.}. For example,
the shape of a soap bubble minimizes its surface, the path followed by
a light ray minimizes (or sometimes maximizes) its travel time
(Fermat principle of least time). It happens that all the fundamental
equations of physics can be derived from such an optimization
principle (in classical mechanics, we talk of a principle of least
action).  By global we mean that the quantity to be optimized depends
on a whole set of points (it is an integral) like the total length of
a path. Noether's theorem says that invariance of the rules used to
compute such global quantity implies the existence of a local
conserved quantity, i.e. computed from some quantities attached to
 one point (and its immediate neighbourhood) of the solution 
 (the orientation of the tangent vector
at each point of the shortest path for instance).

As far as I know, all the conservation laws can be seen as
consequences of Noether's theorem.

(c1)~The invariance under the time shifts implies the existence of
a conserved quantity called energy.

In a closed system, energy cannot be neither lost neither
created\footnote{According to
the most famous formula, the mass is just a form of energy
and therefore the Lavoisier principle of conservation of mass 
can be overruled, but this requires a large amount of energy.}. There are only conversions (from radiation to electricity,
from nuclear energy to thermal energy, from chemical energy to
mechanical energy and so on).  It was because of an apparent small
lack of energy in the~$\beta$-disintegration, that the existence of
the neutrino was inferred by Pauli in 1930. This particle interacts so
weakly with the matter that its energy was missed by the detectors.

(c2) The invariance under the space translations implies 
conservation of the linear momentum (mass$\times$velocity).

This law is manifest both in
the recoil of a canon shooting a bullet and in the recoil of
an atom emitting a photon and rules also the collisions on a billiard or
in the Large Hadron Collider in CERN.  

(c3) The invariance under the space rotations implies 
conservation of the angular momentum (mass$\times$velocity$\times$distance to the rotation center).

Here we take advantage of the isotropy of space: no absolute
direction is preferred. These conservation laws govern the way an ice 
skater controls the rotation speed of her pirouette, imposes helicopters to 
have a rotor mounted on the tailboom to compensate the changes of angular momentum of the main rotor, makes residue of exploded stars 
rotate very quickly and form pulsars.

(c4) The invariance under the some specific gauge transformation implies 
conservation of the electric charge.

This conservation prevents the disintegration of a neutron (non
charged) into a proton plus a photon (non charged): here, the charge
carried by the proton cannot come out of the blue. 

It is worth to notice that such conservation laws are found in the
tiniest accessible corners of our world as well as in the largest
scales of our universe. Noether's theorem allows to understand this
universality: it is directly linked with one of the keystones of
science, the reproducibility.

 Most of the quantum properties are generally extremely 
fragile and are lost in any measurement process.
Quantum particles, even those we consider to be elementary,
can be destroyed or created in reactions. 
 In fact, the conserved quantities 
are the only stable attributes on which one can rely 
and all the quantum particles appear to be a manifestation of
such relatively stable properties.
\begin{quotation}
Elementary particles embody the symmetries; they are their simplest
presentations, and yet they are merely their
consequence.\\ \citep[chap~XX, p.~240]{Heisenberg71a}
\end{quotation}

\section{Beyond physics (but without metaphysics): forging concepts and rational thinking}\label{sec:philosophy}

As remarked by Werner Heisenberg, the conservation laws reactivate the
``problem of change'' (or of becoming)  already been settled by
pre-Athenian philosophers among which Heraclitus and Parmenides, who
proposed the two most extreme solutions along the whole continuous
spectrum of possible answers.
\begin{quotation}For our senses the world consists of an
infinite variety of things and events, colors and sounds. But in order
to understand it we have to introduce some kind of order, and order
means to recognize what is equal, it means some sort of unity. From
this springs the belief that there is one fundamental principle, and
at the same time the difficulty to derive from it the infinite variety
of things.[\dots] This leads to the antithesis of Being and Becoming
and finally to the solution of Heraclitus, that the change itself is
the fundamental principle; the `imperishable change, that renovates
the world,' as the poets have called it. But the change in itself is
not a material cause and therefore is represented in the philosophy of
Heraclitus by the fire as the basic element, which is both matter and
a moving force.\\ \indent We may remark at this point that modern
physics is in some way extremely near to the doctrines of
Heraclitus. If we replace the word `fire' by the word `energy' we can
almost repeat his statements word for word from our modern point of
view. Energy is in fact the substance from which all elementary
particles, all atoms and therefore all things are made, and energy is
that which moves. Energy is a substance, since its total amount does
not change, and the elementary particles can actually be made from
this substance as is seen in many experiments on the creation of
elementary particles. Energy can be changed into motion, into heat,
into light and into tension. Energy may be called the fundamental
cause for all change in the world.  \\ \citep[chap.~IV]{Heisenberg58a}
\end{quotation}
\begin{figure}[!Ht]\begin{center}
\includegraphics[width=8cm]{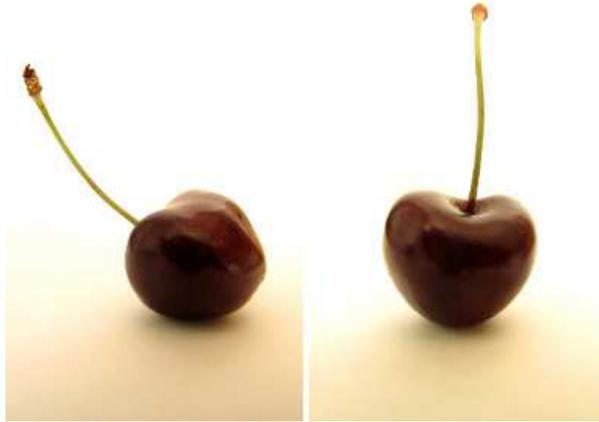}\caption{\label{fig:cerises} What we call ``this cherry'' is an equivalence class, i.e. a set of
 properties like the shape or the color, that remain invariant under an 
uncountable numbers of transformations of irrelevant parameters like 
the rotation of its body, some bending of its tail, the orientation
 of the observer, the intensity of the lighting, the position of the moons of Jupiter, etc. 
 Any concept or fragment of reality requires extraction (abstraction) of some stable properties (at
least for a moment long enough to be noticeable).
During this process, the physical information has been reduced (simplification) and comes with some validity domain that become fuzzy if too much precision is required (at the nanoscale, the boundary of the cherry is not well defined
because, for instance, of the perpetual absorption and desorption of molecules).
}
\end{center}
\end{figure}
After all, what Heisenberg wrote for quantum particles (see the end of
\S~\ref{sec:noether}) can be transposed to any more familiar object
or concept. No thoughts, no language could be possible if we were
unable to discern some ``constant bond'' as Poincar\'e calls it in the
quotation that opens the present work.  These stable properties (at
least for a moment long enough to be noticeable) reveal what we call
the existence of things, building up together what we call the \textit{real
world} (including ourselves). A particular
macroscopic object like a cherry (figure~\ref{fig:cerises}) can be
considered as the result of a \emph{simplification} process, an
\emph{abstraction} in the etymological sense of the word, where a
small cluster of relevant properties have been pruned out from a bunch
of irrelevant phenomena. Its shape, taste, colour, size, etc. remain
invariant despite the change of light, the variations of the point of view,
 its rotations, the endless
adsorption or desorption of molecules. 

With this necessary research of simplicity we come back to the aesthetical  primary signification of
the word ``symmetry'': an elegant efficiency.

\begin{quotation} Like a pure sound or a melodic system of pure sounds in the midst of
noises, so a \emph{crystal}, a \emph{flower}, a \emph{sea shell}
 stand out from the common
disorder of perceptible things. For us they are privileged objects,
more intelligible to the view, although more mysterious upon
reflection, than all those which we see indiscriminately. They present
us with a strange union of ideas: order and fantasy, invention and
necessity, law and exception. In their appearance we find a kind of
\emph{intention} and \emph{action} that seem to have fashioned
them rather as man might have done, but as the same time we find
evidence of methods forbidden and inaccessible to us. We can imitate
these singular forms; our hands can cut a prism, fashion an imitation
flower, turn or model a shell; we are even able to express their
characteristics of symmetry in a formula, or represent them quite
accurately in a geometric construction. Up to this point we can share
with ``nature'': we can endow her with designs, a sort of mathematics,
a certain taste and imagination that are not infinitely different from
ours; but then, after we have endowed her with all the human qualities
she needs to make herself understood by human beings, she displays all
the inhuman qualities needed to disconcert us\dots\\
\citep{Valery64a}.
\end{quotation}


\begin{thebibliography}{214}
\expandafter\ifx\csname natexlab\endcsname\relax\def\natexlab#1{#1}\fi
\providecommand{\url}[1]{\texttt{#1}}
\providecommand{\href}[2]{#2}
\providecommand{\path}[1]{#1}
\providecommand{\DOIprefix}{doi:}
\providecommand{\ArXivprefix}{arXiv:}
\providecommand{\URLprefix}{}
\providecommand{\Pubmedprefix}{pmid:}
\providecommand{\doi}[1]{\href{http://dx.doi.org/#1}{\path{#1}}}
\providecommand{\Pubmed}[1]{\href{pmid:#1}{\path{#1}}}
\providecommand{\bibinfo}[2]{#2}
\ifx\xfnm\relax \def\xfnm[#1]{\unskip,\space#1}\fi

\bibitem[{Heisenberg(1958)}]{Heisenberg58a}
\bibinfo{author}{Heisenberg, W.}, \bibinfo{year}{1958}.
\newblock \bibinfo{title}{Physics and philosophy. The revolution in modern
  science}. volume \bibinfo{volume}{1958} of \textit{\bibinfo{series}{World
  perspectives}}.
\newblock \bibinfo{publisher}{Harper and brothers}, \bibinfo{address}{New York}.
\bibitem[{Heisenberg(1971)}]{Heisenberg71a}
\bibinfo{author}{Heisenberg, W.}, \bibinfo{year}{1971}.
\newblock \bibinfo{title}{Physics and beyond --- Encounters and Conversations}.
\newblock \bibinfo{publisher}{Harper and Row}, \bibinfo{address}{New  York}.
\newblock \bibinfo{note}{Translated by Arnold. J. Pomerans from the German original edition \textit{Der Teil  und das  Ganze --- Gespr{\"a}che im  Umkreis der Atomphysik} (Piper et co. Verlag, Munich, 1969)}.
\bibitem[{{K}osmann{-S}chwarzbach(2010)}]{KosmannSchwarzbach10a}
\bibinfo{author}{{K}osmann{-S}chwarzbach, Y.}, \bibinfo{year}{2010}.
\newblock \bibinfo{title}{The {N}oether theorems. Invariance and conservation
  laws in the twentieth century (sources and studies in the history of
  mathematics and physical sciences)}.
\newblock \bibinfo{publisher}{Springer}.
\newblock \bibinfo{note}{Translated by Bertram E. Schwarzbach from the \ French
  original edition \textit{Les Th\'eor\`emes de Noether. Invariance et lois de conservation au XXe si\`ecle}, Les {\'E}ditions de {l'\'E}cole Polytechnique,2004}.
\bibitem[{Mouchet(2013a)}]{Mouchet13a}
\bibinfo{author}{Mouchet, A.}, \bibinfo{year}{2013}a.
\newblock \bibinfo{title}{L'{\'e}l{\'e}gante efficacit{\'e} des sym{\'e}tries}.
\newblock UniverSciences, \bibinfo{publisher}{Dunod}, \bibinfo{address}{Paris}.
\bibitem[{Mouchet(2013b)}]{Mouchet13b}
\bibinfo{author}{Mouchet, A.}, \bibinfo{year}{2013}b.
\newblock \bibinfo{title}{Reflections on the four facets of symmetry: 
how physics exemplifies rational thinking}.
\newblock \bibinfo{journal}{Eur. Phys. J. H} \bibinfo{volume}{38},
  \bibinfo{pages}{661--702}.
\bibitem[{Poincar{\'e}(1895)}]{Poincare1895a}
\bibinfo{author}{Poincar{\'e}, H.}, \bibinfo{year}{1895}.
\newblock \bibinfo{title}{L'espace et la g\'eom\'etrie}.
\newblock \bibinfo{journal}{Revue de m{\'e}taphysique et de morale}
  \bibinfo{volume}{3}, \bibinfo{pages}{631--646}.
\newblock \bibinfo{note}{Reproduced with modifications in chap.~IV of
  \cite{Poincare52a}}.
\bibitem[{Poincar{\'e}(1902)}]{Poincare02a}
\bibinfo{author}{Poincar{\'e}, H.}, \bibinfo{year}{1902}.
\newblock \bibinfo{title}{Sur la valeur objective de la science}.
\newblock \bibinfo{journal}{Revue de m{\'e}taphysique et de morale}
  \bibinfo{volume}{10}, \bibinfo{pages}{263--293}.
\newblock \bibinfo{note}{Reproduced as chap.~X and XI of \cite{Poincare58a}}.
\bibitem[{Poincar{\'e}(1952)}]{Poincare52a}
\bibinfo{author}{Poincar{\'e}, H.}, \bibinfo{year}{1952}.
\newblock \bibinfo{title}{Science and hypothesis}.
\newblock \bibinfo{publisher}{{D}over {P}ublications, {I}nc.},
  \bibinfo{address}{{N}ew {Y}ork}.
\newblock \bibinfo{note}{Translated by W.~J.~Greenstreet from the French original edition
  \textit{La science et l'hypoth\`ese} (Flammarion, 1902)}.
\bibitem[{Poincar{\'e}(1958)}]{Poincare58a}
\bibinfo{author}{Poincar{\'e}, H.}, \bibinfo{year}{1958}.
\newblock \bibinfo{title}{The Value of Science}.
\newblock \bibinfo{publisher}{{D}over {P}ublications, {I}nc.},
  \bibinfo{address}{{N}ew {Y}ork}.
\newblock \bibinfo{note}{Translated by George Bruce Halsted from the French original edition
  \textit{La valeur de la science} (Flammarion, 1905)}.
\bibitem[{Val\'ery(1964)}]{Valery64a}
\bibinfo{author}{Val\'ery, P.}, \bibinfo{year}{1964}.
\newblock \bibinfo{title}{Aesthetics}.
\newblock Bollinger Series XLV, vol.~13, \bibinfo{publisher}{Pantheon Books},
  \bibinfo{address}{New York}.
\newblock \bibinfo{note}{\textit{Man and the Sea Shell} translated by Ralph
  Manheim from the French \textit{L'homme et la coquille} (Vari\'et\'e V,
  Gallimard)}.
\bibitem[{Yourcenar(1976)}]{Yourcenar76a}
\bibinfo{author}{Yourcenar, M.}, \bibinfo{year}{1976}.
\newblock \bibinfo{title}{The abyss}.
\newblock \bibinfo{publisher}{Farrar, Straus and Giroux}, \bibinfo{address}{New
  York}.
\newblock \bibinfo{note}{English translation by Grace Frick in collaboration
  with the author from the French original edition \textit{L'{\oe}uvre au noir}
  (Gallimard, 1968).}
\end{thebibliography}
\end{document}